\documentclass[apl,amsmath,amssymb,reprint,superscriptaddress]{revtex4-2}

\usepackage{amssymb}
\usepackage{lipsum}
\usepackage{float}
\usepackage{subfig}
\usepackage{caption}
\usepackage{subcaption}
\usepackage{graphicx}
\usepackage{dcolumn}
\usepackage{bm}

\usepackage{siunitx}
\usepackage{upgreek}
\usepackage[utf8]{inputenc}
\usepackage[T1]{fontenc}
\usepackage{mathptmx}
\usepackage{etoolbox}
\usepackage{pdfpages}
\usepackage{pgffor}
\makeatletter
\AtBeginDocument{\let\LS@rot\@undefined}
\makeatother

\def\supplementfilename{SI.pdf}

\pdfximage{\supplementfilename}
\def\numbersupplementpages{\the\pdflastximagepages}

\newif\ifarXiv
\arXivtrue 

\begin{document}


\title{The impact of electrical contacts on the optical properties of a  MoS$_{2}$ monolayer} 



\author{Agata Zielińska}
\email[]{agata.zielinska@pwr.edu.pl}
\affiliation{Department of Experimental Physics, Wrocław University of Science and Technology, Wybrzeże Wyspiańskiego 27, 50-370, Wrocław, Poland}
\author{Mateusz Dyksik}
\affiliation{Department of Experimental Physics, Wrocław University of Science and Technology, Wybrzeże Wyspiańskiego 27, 50-370, Wrocław, Poland}
\author{Alessandro Surrente}
\affiliation{Department of Experimental Physics, Wrocław University of Science and Technology, Wybrzeże Wyspiańskiego 27, 50-370, Wrocław, Poland}
\author{Jonathan Eroms}
\affiliation{Institute for Experimental and Applied Physics, University of Regensburg, Universitätsstraße 31, 93053, Regensburg, Germany}
\author{Dieter Weiss}
\affiliation{Institute for Experimental and Applied Physics, University of Regensburg, Universitätsstraße 31, 93053, Regensburg, Germany}
\author{Paulina Płochocka}
\affiliation{Department of Experimental Physics, Wrocław University of Science and Technology, Wybrzeże Wyspiańskiego 27, 50-370, Wrocław, Poland}
\affiliation{Laboratoire National des Champs Magnétiques Intenses, 143 avenue de Rangueil, 31400, Tolouse, France}
\author{Mariusz Ciorga}
\affiliation{Department of Experimental Physics, Wrocław University of Science and Technology, Wybrzeże Wyspiańskiego 27, 50-370, Wrocław, Poland}


\date{\today}

\begin{abstract}
Achieving high performance in transition-metal-dichalcogenide-based optoelectronic devices is challenging -- the realization of an efficient electrical contacting scheme should not be obtained at the expense of their optical quality. Here we present the optical properties of MoS$_{2}$ monolayers which have been electrically contacted with bismuth and gold. The photoluminescence (PL) spectrum of the samples contacted with both materials is significantly broadened. In the case of the bismuth contacted sample we note an additional, low energy band in the PL spectrum, attributed to a defect state formed during the evaporation of Bi. Comparing the intensity of the excitonic peak and of the defect-related peak, we note that there is a correlation between the type of contacts and the optical properties.
\end{abstract}


\maketitle 

The unique electronic and optical properties of semiconducting van der Waals (vdW) materials make them attractive and promising candidates for optoelectronic devices \cite{Liu2019,Avsar2020, mueller_exciton_2018,Zollner2019}. Among the most extensively studied two-dimensional (2D) materials are the transition metal dichalcogenides (TMDCs), in particular MoS$_{2}$, MoSe$_{2}$, WS$_{2}$ and WSe$_{2}$ \cite{WangBook,KolobovBook,WeeBook}. These materials exhibit strongly bound excitons with high oscillator strength \cite{Chernikov2014,Back2018}, which led naturally to their study by means of various optical spectroscopic techniques. Excellent optical properties of these materials are also followed by interesting electrical properties, such as the valley Hall effect \cite{Mak2014,Habe2017}. 

An efficient contacting scheme is essential in order to fully exploit the optoelectronic potential of these materials and this has been proven to be challenging in the case of TMDCs \cite{schulman2018,allain2015,zheng2021}. Recent reports demonstrate the possibility of obtaining good ohmic contacts between a semimetallic material (bismuth or antimony) and a semiconducting TMDC layer \cite{shen2021,li_approaching_2023,Lee2024}. Due to the near-zero density of states at the Fermi level of the semimetal, the formation of gap states in TMDCs, which are responsible for the Fermi-level pinning at the metal-TMDC interface, is sufficiently suppressed, making it possible to obtain low-resistance ohmic contacts to the TMDC. In order to fully exploit the potential of semimetal contacts to TMDCs, the exact physics at a semimetal/TMDC interface needs to be explored and understood \cite{Su2023}. However, combining optical and electrical investigations on the same sample is very challenging, due to the numerous processing steps needed for efficient contacting, which can lead to degradation of the optical quality of the sample \cite{Turunen2023}. Furthermore, the materials of choice for the fabrication of electrical contacts are usually not transparent to visible or infrared light \cite{Fan2018,Johnson1972}. Finally, the optical properties of the sample can be influenced by the charge transfer through the contacts \cite{Ahmed2020,Grzeszczyk2020}.

In this work, we extend the approach of bismuth contacting by monitoring the emission response of MoS$_{2}$/Bi heterostructures. We aim to investigate the influence of bismuth contacts on the optical properties of a MoS$_{2}$ monolayer and to establish a correlation between the type and quality of the contacts and the optical properties of MoS$_{2}$. With this goal in mind, we prepared a series of MoS$_2$ monolayer samples with bismuth as the contact material. For all investigated samples we measured the electrical performance and the optical properties of the MoS$_{2}$ flakes. Additionally, as a reference, we prepared and measured non-contacted and Au-contacted MoS$_{2}$ flakes.

\begin{figure*}[ht]
    \centering
    \includegraphics[width=\textwidth]{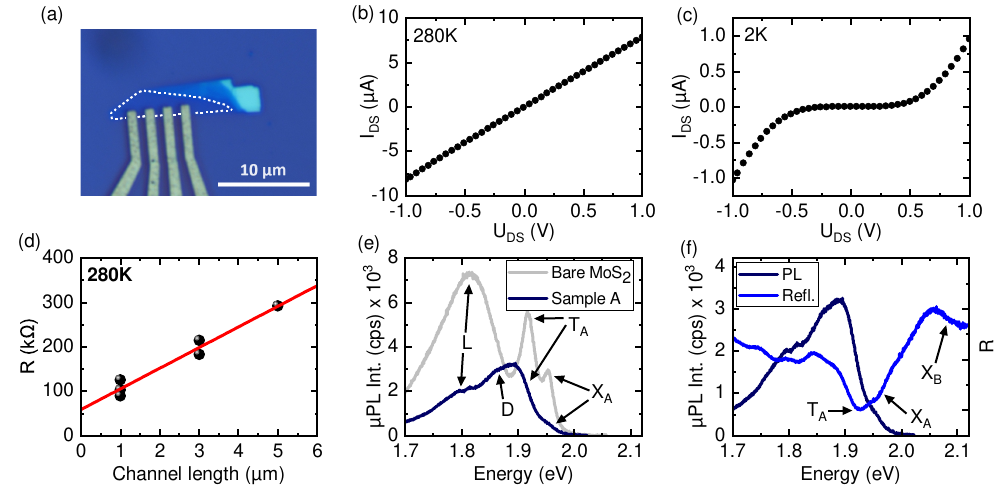}
    \caption{(a) Optical micrograph of sample A. All contacts are \SI{1}{\micro\metre} wide and \SI{100}{\nano\metre} high, with \SI{1}{\micro\metre} spacing between them. MoS$_{2}$ monolayer (marked with the white dotted line) is placed below the contacts. (b) I-V curve for sample A for zero backgate voltage at \SI{280}{\kelvin}. The contacts show clear ohmic behaviour. (c) I-V curve for sample A at \SI{2}{\kelvin}. The ohmic behaviour is lost but some conduction through the sample can still be observed. (d) Measured total resistance as a function of MoS$_{2}$ channel length at \SI{280}{\kelvin} for sample A. The total resistance increases linearly with the channel length. (e) $\upmu$PL spectrum of sample A at \SI{4.5}{\kelvin} (dark blue) compared to a representative spectrum of the bare MoS$_{2}$ flake (gray). Defect-related emission (L), new emission peak (D), trion emission (T$_{\text{A}}$) and exciton emission (X$_{\text{A}}$) are marked with arrows. (f) $\upmu$PL (dark blue) and reflectivity (blue) spectra. In the reflectivity spectrum, two transitions are well resolved: $\sim \SI{1.923}{\eV}$ (corresponding to T$_{\text{A}}$) and $\sim \SI{2.085}{\eV}$ (corresponding to X$_{\text{B}}$). The transition at $\sim \SI{1.956}{\eV}$, corresponding to X$_{\text{A}}$, is weakly visible.}
    \label{fig1}
\end{figure*}

We first discuss the optoelectronic properties of the representative sample of the back-gated MoS$_2$ monolayers with bismuth contacts, which we will refer to as sample A. We show its micrograph in Fig.\ \ref{fig1}(a). The layer was placed on a Si/SiO$_{2}$ substrate, and bismuth contacts were evaporated in an ultra-high vacuum (UHV) system using an electron gun. We also prepared a series of reference samples with gold as a contact material. See the Supplementary Material for more details on the fabrication process.

We measured the current-voltage (I-V) curves of sample A for temperatures of \SIrange{2}{280} {\kelvin}. Fig.\ \ref{fig1}(b) shows the representative I-V curve for zero back-gate voltage at \SI{280}{\kelvin}. The curve shows ohmic behavior, confirming that bismuth is a good contact material for MoS$_{2}$ at room temperature \cite{shen2021,li_approaching_2023,Lee2024}. The contacts lose their ohmic characteristics at temperatures below \SI{80}{\kelvin} (see Fig.\ \ref{fig1}(c)), but they still conduct current. Fig.\ \ref{fig1}(d) presents the total resistance as a function of channel length (distance between contacts). We can observe the resistance increasing linearly with the channel length, which indicates that the total resistance is dominated by the resistance of the channel.

Sample A was optically investigated by measuring microreflectivity and microphotoluminescence ($\upmu$PL) spectra at \SI{4.5}{\kelvin}. Fig.\ \ref{fig1}(e) shows a representative PL spectrum of sample A (dark blue line) and of the bare MoS$_{2}$ monolayer exfoliated directly on the SiO$_2$ substrate (gray line). The bare flake exhibits the typical low temperature PL spectrum of MoS$_{2}$, with three distinct emission lines. We attribute the high energy peak at $\sim \SI{1.95}{\eV}$ to the recombination of the neutral exciton (X$_{\text{A}}$), the peak at $\sim \SI{1.92}{\eV}$ to the recombination of the charged exciton (T$_{\text{A}}$), while the low-energy broad band, labelled L in Fig.\ \ref{fig1}(e), is related to the recombination of excitons bound to defects present in the MoS$_2$ monolayer \cite{Jadczak2017, Munson2024}. The higher intensity of the T$_{\text{A}}$ line compared to that of the X$_{\text{A}}$ line indicates charge doping, which in the case of MoS$_{2}$ has an n-type character, as MoS$_{2}$ is a naturally n-doped material \cite{Park2023}. This is consistent with the observation of n-type current in electrical measurements. The PL spectrum of sample A is significantly different from that of the bare flake. The total intensity of the observed emission is strongly quenched and the emission lines are not so well resolved anymore. The maximum of the emission is now at $\sim\SI{1.9}{\eV}$, i.e., at slightly lower energy than the T$_{\text{A}}$ line. This is the result of a new, broad emission line appearing at $\sim\SI{1.86}{\eV}$, i.e., between the T$_{\text{A}}$ and L lines, which we label D. To properly assign the observed emission lines to the corresponding optical transitions we performed reflectivity measurements on the investigated samples. In Fig.\ \ref{fig1}(f) we compare the PL spectrum of sample A with a reflectivity spectrum measured on another MoS$_{2}$ monolayer sample with bismuth contacts. The transitions clearly visible in the reflectivity spectrum correspond to the charged exciton at \SI{1.92}{\eV}, the A exciton at \SI{1.96}{\eV}, and the B exciton at \SI{2.1}{\eV} \cite{WangBook}. The presence of a trion-related resonance in the reflectivity spectrum suggests significant doping, which induces a partial transfer of the oscillator strength from the exciton to the charged exciton transition \cite{Courtade2017,Sidler2017}. All exciton transitions in reflectivity are in good agreement with the literature values \cite{tran2016}. However, there is no visible transition in the reflectivity at $\sim\SI{1.86}{\eV}$, which suggests that line D corresponds to the defect-related emission (discussed in the following).

\begin{figure*}[!htbp]
    \centering
    \includegraphics[width=\textwidth]{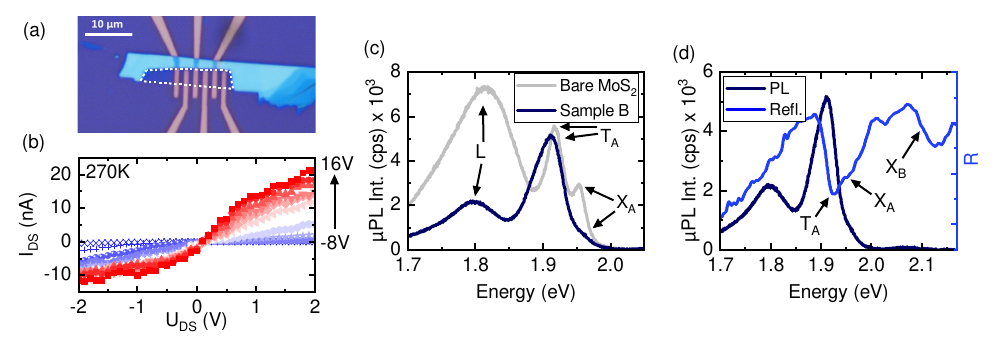}
    \caption{(a) Optical microscope image of sample B. All contacts are \SI{25}{\nano\metre} high. The MoS$_{2}$ monolayer (marked with the white dotted line) is placed on top of the contacts. Adapted from Ref. \cite{Zielinska2025}. (b) Representative I-V curves for sample B for back gate voltages in the range \SIrange{-8}{16}{\volt} measured at \SI{270}{\kelvin}. The shape of the curves suggests a Schottky-type contacts, which are also sensitive to the back gate voltage. Adapted from Ref. \cite{Zielinska2025}. (c) $\upmu$PL spectrum of sample B (dark blue) compared to the $\upmu$PL spectrum of the bare MoS$_{2}$ flake (gray) measured at \SI{4.5}{\kelvin}. Defect-related emission (L), trion emission (T$_{\text{A}}$) and exciton emission (X$_{\text{A}}$) are marked with arrows. (d) $\upmu$PL (dark blue) and reflectivity (blue) spectra of sample B. In the reflectivity spectrum two transitions are well resolved: $\sim \SI{1.92}{\eV}$ and $\sim \SI{2.1}{\eV}$. The transition at $\sim \SI{1.95}{\eV}$ is weakly visible.}
    \label{fig2}
\end{figure*}
To check whether the D line is characteristic only for the Bi-contacted samples, we also prepared and measured a set of samples with Au contacts \cite{Zielinska2025}, and the optical microscope image of the representative Au-contacted sample (sample B) is presented in Fig.\ \ref{fig2}(a). Whereas more details on fabrication, processing, and results of optical and electrical measurements of these samples can be found in the Supplementary Material, here we present the most relevant features. In Fig. \ref{fig2}(b) we show the representative I-V curves measured at \SI{270}{\kelvin}, showing Schottky-like characteristics. In Fig. \ref{fig2}(c) and (d) we show, respectively, a representative PL and reflectance spectra of sample B at \SI{4.5}{\kelvin}. Similarly as for sample A, the PL lines are broadened compared to those for the bare MoS$_{2}$, but there is no signature of any additional line close to the $T_A$  line. This leads us to the conclusion that the apparent additional line labelled D in Fig.\ \ref{fig1} appears as a result of bringing the MoS$_2$ flake into contact with Bi. In similar experiments performed on CVD-grown MoS$_2$ monolayers covered with thin semimetal layers \cite{feng_direct_2024}, Feng \textit{et al.} observed red shift of the PL emission at room temperature. They explained it as a consequence of the transfer of electrons from Bi to MoS$_2$, resulting in an increase in the electron concentration in MoS$_2$ and an increase of the intensity of the T$_{\text{A}}$ emission relative to the X$_{\text{A}}$ emission. This transfer of electrons from Bi to MoS$_2$ is not unexpected, as theoretical calculations show that at the Bi / MoS$_2$ interfaces, electrons tend to accumulate at the S atoms due to their stronger electronegativity than for the Bi atoms \cite{Su2023,feng_direct_2024}. One cannot exclude  possible diffusion of Bi atoms into MoS$_2$, which would result in electron doping; however, there is not much information in the literature on doping of MoS$_2$ with Bi \cite{sovizi_single_2022}. Qumar \textit{et al.} \cite{Qumar2020} observed a small redshift in PL from Bi-doped MoS$_2$ nanosheets with Bi atoms; however, they did not explicitly state whether Bi substituted S or Mo atoms. The diffusion of Bi atoms into MoS$_{2}$ would also be consistent with the fact that the D line appears on the entire surface of the sample, also on parts of the flake far from the contacts. 

\begin{figure*}[htbp!]
    \centering
    \includegraphics[width=\textwidth]{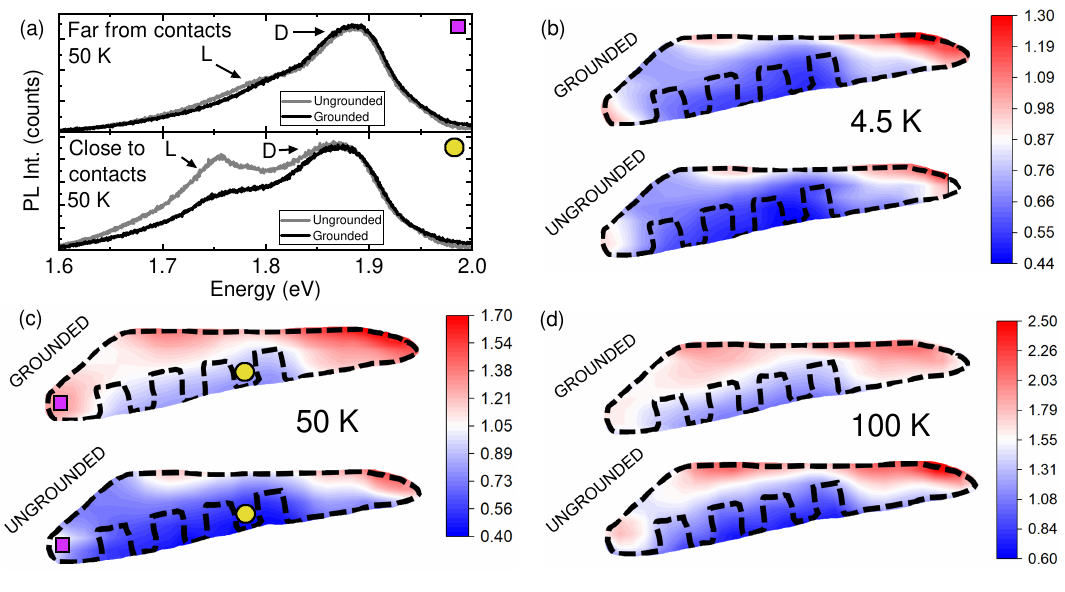}
    \caption{(a) Comparison of PL spectra of two different spots on the sample A for contacts grounded and ungrounded at \SI{50}{\kelvin}. The violet square corresponds to the spot far from the contacts and the yellow circle corresponds to the spot close to the contacts (both spots are marked on the corresponding map below). (b)-(d) Trion-to-defect ratio for the contacts grounded and ungrounded at \SI{4.5}{\kelvin}, \SI{50}{\kelvin} and \SI{100}{\kelvin} for sample A. The dashed lines represent the edges of the flake and the contour of the contacts. For \SI{50}{\kelvin} a substantial difference is visible in the grounded contacts as compared to the ungrounded contacts. In case of \SI{4.5}{\kelvin} and \SI{100}{\kelvin} we observe smaller differences in the map, however, the trion to defect ratio is also bigger for the grounded contacts.}
    \label{fig3}
\end{figure*}

To check how the presence of excess free carriers in the flake affects the PL  spectrum, we performed $\upmu$PL measurements at \SI{4.5}{\kelvin}, \SI{50}{\kelvin}, \SI{100}{\kelvin} and \SI{300}{\kelvin} with all electrical contacts grounded and not grounded, assuming that the grounding removes excess charge carriers. Generally, in case of sample A, there is a substantial difference in the optical spectrum when we ground the contacts or leave them ungrounded. Grounding leads to a decrease of the intensity of defect L and to negligible changes in the intensity of trion and D lines, as illustrated in Fig.\ \ref{fig3} (a), where we plot the PL spectra of two spots on the sample - far from contacts (the violet square) and close to contacts (the yellow circle). We can observe the quench of defect L emission on both spots, but in case of the spot close to the contacts, this effect is more pronounced.  Fig.\ \ref{fig3}(b)-(d) shows $\upmu$PL maps of the trion-to-defect ratio at selected temperatures for grounded and ungrounded contacts for sample A, where the intensity ratio was obtained by integration in the ranges corresponding to the trion at $\sim$\SI{1.9}{\eV} (\SI{0.14}{\eV} range width) and to the defect-bound excitons at $\sim$\SI{1.8}{\eV} (\SI{0.19}{\eV} range width), where the former range includes also the line D at \SI{1.86}{\eV}. The ranges were red-shifted with the temperature, according to the temperature shift of the peaks. The ratio is clearly larger for grounded contacts, indicating that excess charge carriers transferred across the contacts bind mainly to the defects present in the crystal, while the trion and exciton emission lines are not significantly influenced. The effect is the strongest at \SI{50}{\kelvin}. We relate this to the fact that at this temperature the contacts have much smaller resistance compared to the resistance at \SI{4.5}{\kelvin}, which increases the possibility of charge transfer. In addition, at low temperatures, the carriers are tightly bound, which means they can not be freely transferred through the contacts. However, the defect emission is generally significantly suppressed at higher temperatures, which would make comparison with the trion intensity impossible. We can conclude that the charge transfer through the working contacts mostly affects the excess charge in the flake. As D emission is not significantly influenced by grounding, we can conclude that it is of a different nature than L. 

\begin{figure}
    \centering
    \includegraphics[width=\columnwidth]{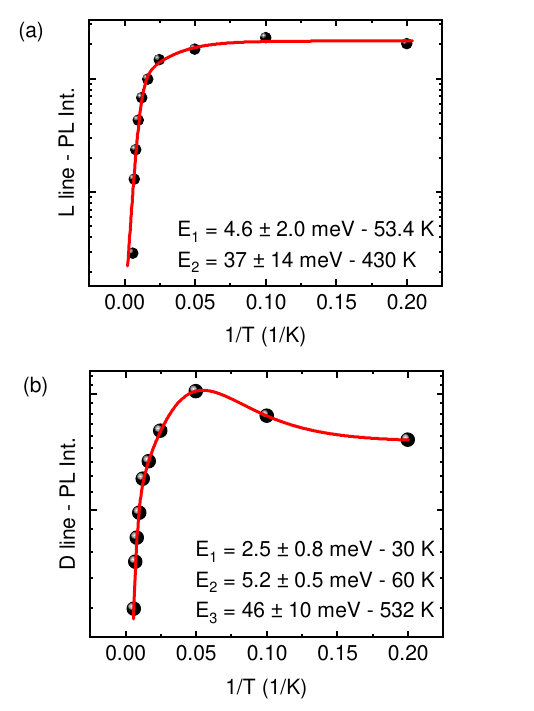}
    \caption{The Arrhenius plot of peak L (a) and D (b) intensity with the Arrhenius equations fitted to the data.}
    \label{fig4}
\end{figure}

To analyze in more depth the effects described above, we measured the temperature dependence of the PL spectrum of sample A in the range of \SIrange{5}{300}{\kelvin}. We fitted the curves with four Gaussian peaks to obtain the emission intensity of the L and D lines as a function of temperature and fitted the Arrhenius plot of the obtained data to extract the activation energies of the relevant processes of exciton dissociation \cite{Shibata_1998}. In case of line L we used the formula with two activation energies for PL quenching:
\begin{equation}
    I=\dfrac{I_{0}}{1+A_{1}\exp (-E_{1}/k_{B}T)+A_{2}\exp (-E_{2}/k_{B}T)}.
\end{equation}
In case of peak D we used the formula with three activation energies - one for the increase of the PL at low temperatures and two for PL quenching at higher temperatures:
\begin{equation}
    I=\dfrac{I_{0}[1+(A_{1}\exp (-E_{1}/k_{B}T)]}{1+A_{2}\exp (-E_{2}/k_{B}T)+A_{3}\exp (-E_{3}/k_{B}T)},
\end{equation}
where $E_{i}$ is the activation energy and $A_{i}$ is the amplitude of the process. The results of the fitting are presented in Fig. \ref{fig4}. We compared the obtained $E_{i}$ values with the binding energies of various excitonic complexes on different types of defects in TMDCs, calculated with Monte Carlo simulations by Mostaani \textit{et al.} \cite{Mostaani2017}. In case of line L, the lower activation energy of \SI{4.6}{\milli\eV} is consistent, within the measurement error, with the $A^-X$ dissociation, with the acceptor states possibly created by a carbon atom placed in a sulphur vacancy. This energy corresponds to a thermal energy at \SI{53.4}{\kelvin}, which means that at $\sim$\SI{50}{\kelvin} the carriers are not tightly bound and some of them can escape through the potential barrier at the contacts. The higher energy of $\sim$ \SI{37}{\milli\eV} could indicate dissociation of  a $D^0X$ complex. In case of line D,  the activation energy of  \SI{2.5}{\milli\eV} for the initial increase in PL is also consistent  with the dissociation of $A^-X$. This could mean that the carriers previously dissociated from the line L bind to the states responsible for D emission, although another, unknown reservoir of carriers could also be involved. The activation energy of the first quench process of line D, equal to \SI{5.2}{\milli\eV}, may be related to the dissociation of the $D^{+}X$ or $A^{-}X$ complex. This energy corresponds to the thermal energy at \SI{60}{\kelvin}. The fact that despite that, the line D is not significantly influenced by grounding at \SI{50}{\kelvin} could be explained by the supply of carriers from the above-mentioned  reservoir with the activation energy of \SI{2.5}{\milli\eV}, which could replace the carrier that escape through the potential barrier at the contacts.  The second quenching process, with an activation energy of \SI{46}{\milli\eV}, is consistent with the dissociation of a biexciton bound to an acceptor ($A^-XX$), with a calculated $E_{A}=$  \SI{48}{\milli\eV}.  Therefore, if we assume the diffusion of Bi atoms during the evaporation process and formation of acceptor states as a result of sulphur substitution, we could explain the line D as originating from the emission of excitons and biexcitons bound to Bi acceptors. A more detailed analysis of the Arrhenius plots is presented in the Supplementary Material. There, we also show the trion-to-defect maps of sample B at \SI{50}{\kelvin} with grounded and ungrounded contacts. There is no significant difference between the two maps, which probably results from the high resistance of the Schottky contacts at low temperatures. 

In conclusion, we presented the properties of MoS$_{2}$ monolayers, which have been electrically contacted with semimetallic bismuth. The contacts show ohmic behavior in the temperature range from \SI{280}{\kelvin} to \SI{80}{\kelvin}, whereas the gold contacts used as a reference exhibit Schottky characteristics throughout the entire range. The two materials used to fabricate the contacts also significantly affect the optical properties of the MoS$_2$ monolayers. In the case of the bismuth contacts, an additional defect band in the PL spectrum is attributed to the diffusion of bismuth atoms during the deposition process. 
Finally, by comparing the PL spectra of grounded and ungrounded samples at \SI{50}{\kelvin}, we demonstrate that the grounded bismuth contacts significantly reduce the intensity of defect-bound excitons, which we attribute to charge transfer through the grounded contacts removing the charges bound to defects. This effect is not observed in the sample with gold contacts, which is associated with their Schottky-like characteristics. Overall, our results show that bismuth contacts are very strong candidates for electrical contacting in high quality, TMD-based optoelectronic devices. However, their effect on the optical spectrum of TMDC requires further exploration.\\

See the Supplementary Material for details on the processing of the samples, measurements details, and additional results.\\

This work was financially supported from the project no.\ 2022/45/B/ST5/04292, funded by the National Science Centre in Poland. We also acknowledge financial support from the European Union (EU) student exchange programme Erasmus+ Internship.

\section*{Author declaration}
\subsection*{Conflict of Interest}

The authors have no conflicts to disclose.

\subsection*{Author Contributions}
Agata Zielińska: Data Curation (lead); Formal Analysis (lead); Writing - Original Draft Preparation (lead). Mateusz Dyksik: Conceptualization (equal); Methodology (equal); Supervision (supporting); Validation (equal); Writing - Original Draft Preparation (supporting). Alessandro Surrente: Validation (equal); Writing - Original Draft Preparation (supporting). Jonathan Eroms: Resources (equal); Supervision (supporting); Writing - Review and Editing (equal). Dieter Weiss: Resources (equal); Writing - Review and Editing (equal). Paulina Płochocka: Project Administration (supporting); Resources (equal); Validation (equal); Writing - Review and Editing (equal). Mariusz Ciorga: Conceptualization (equal); Funding Acquisition (lead); Methodology (equal); Project Administration (lead); Resources (equal); Supervision (lead); Validation (equal); Writing - Original Draft Preparation (supporting).

\section*{Data availability}

The data that support the findings of this study are available from
the corresponding author upon reasonable request.



%
%

%


\bibliography{aipsamp}

\begin{thebibliography}{35}%
\makeatletter
\providecommand \@ifxundefined [1]{%
 \@ifx{#1\undefined}
}%
\providecommand \@ifnum [1]{%
 \ifnum #1\expandafter \@firstoftwo
 \else \expandafter \@secondoftwo
 \fi
}%
\providecommand \@ifx [1]{%
 \ifx #1\expandafter \@firstoftwo
 \else \expandafter \@secondoftwo
 \fi
}%
\providecommand \natexlab [1]{#1}%
\providecommand \enquote  [1]{``#1''}%
\providecommand \bibnamefont  [1]{#1}%
\providecommand \bibfnamefont [1]{#1}%
\providecommand \citenamefont [1]{#1}%
\providecommand \href@noop [0]{\@secondoftwo}%
\providecommand \href [0]{\begingroup \@sanitize@url \@href}%
\providecommand \@href[1]{\@@startlink{#1}\@@href}%
\providecommand \@@href[1]{\endgroup#1\@@endlink}%
\providecommand \@sanitize@url [0]{\catcode `\\12\catcode `\$12\catcode
  `\&12\catcode `\#12\catcode `\^12\catcode `\_12\catcode `\%12\relax}%
\providecommand \@@startlink[1]{}%
\providecommand \@@endlink[0]{}%
\providecommand \url  [0]{\begingroup\@sanitize@url \@url }%
\providecommand \@url [1]{\endgroup\@href {#1}{\urlprefix }}%
\providecommand \urlprefix  [0]{URL }%
\providecommand \Eprint [0]{\href }%
\providecommand \doibase [0]{https://doi.org/}%
\providecommand \selectlanguage [0]{\@gobble}%
\providecommand \bibinfo  [0]{\@secondoftwo}%
\providecommand \bibfield  [0]{\@secondoftwo}%
\providecommand \translation [1]{[#1]}%
\providecommand \BibitemOpen [0]{}%
\providecommand \bibitemStop [0]{}%
\providecommand \bibitemNoStop [0]{.\EOS\space}%
\providecommand \EOS [0]{\spacefactor3000\relax}%
\providecommand \BibitemShut  [1]{\csname bibitem#1\endcsname}%
\let\auto@bib@innerbib\@empty
\bibitem [{\citenamefont {Liu}\ \emph {et~al.}(2019)\citenamefont {Liu},
  \citenamefont {Gao}, \citenamefont {Zhang}, \citenamefont {He}, \citenamefont
  {Yu},\ and\ \citenamefont {Liu}}]{Liu2019}%
  \BibitemOpen
  \bibfield  {author} {\bibinfo {author} {\bibfnamefont {Y.}~\bibnamefont
  {Liu}}, \bibinfo {author} {\bibfnamefont {Y.}~\bibnamefont {Gao}}, \bibinfo
  {author} {\bibfnamefont {S.}~\bibnamefont {Zhang}}, \bibinfo {author}
  {\bibfnamefont {J.}~\bibnamefont {He}}, \bibinfo {author} {\bibfnamefont
  {J.}~\bibnamefont {Yu}},\ and\ \bibinfo {author} {\bibfnamefont
  {Z.}~\bibnamefont {Liu}},\ }\bibfield  {title} {\bibinfo {title}
  {Valleytronics in transition metal dichalcogenides materials},\ }\href@noop
  {} {\bibfield  {journal} {\bibinfo  {journal} {Nano Research}\ }\textbf
  {\bibinfo {volume} {12}},\ \bibinfo {pages} {2695} (\bibinfo {year}
  {2019})}\BibitemShut {NoStop}%
\bibitem [{\citenamefont {Avsar}\ \emph {et~al.}(2020)\citenamefont {Avsar},
  \citenamefont {Ochoa}, \citenamefont {Guinea}, \citenamefont {\"Ozyilmaz},
  \citenamefont {van Wees},\ and\ \citenamefont {Vera-Marun}}]{Avsar2020}%
  \BibitemOpen
  \bibfield  {author} {\bibinfo {author} {\bibfnamefont {A.}~\bibnamefont
  {Avsar}}, \bibinfo {author} {\bibfnamefont {H.}~\bibnamefont {Ochoa}},
  \bibinfo {author} {\bibfnamefont {F.}~\bibnamefont {Guinea}}, \bibinfo
  {author} {\bibfnamefont {B.}~\bibnamefont {\"Ozyilmaz}}, \bibinfo {author}
  {\bibfnamefont {B.~J.}\ \bibnamefont {van Wees}},\ and\ \bibinfo {author}
  {\bibfnamefont {I.~J.}\ \bibnamefont {Vera-Marun}},\ }\bibfield  {title}
  {\bibinfo {title} {{Colloquium: Spintronics in graphene and other
  two-dimensional materials}},\ }\href@noop {} {\bibfield  {journal} {\bibinfo
  {journal} {Rev. Mod. Phys.}\ }\textbf {\bibinfo {volume} {92}},\ \bibinfo
  {pages} {021003} (\bibinfo {year} {2020})}\BibitemShut {NoStop}%
\bibitem [{\citenamefont {Mueller}\ and\ \citenamefont
  {Malic}(2018)}]{mueller_exciton_2018}%
  \BibitemOpen
  \bibfield  {author} {\bibinfo {author} {\bibfnamefont {T.}~\bibnamefont
  {Mueller}}\ and\ \bibinfo {author} {\bibfnamefont {E.}~\bibnamefont
  {Malic}},\ }\bibfield  {title} {\bibinfo {title} {Exciton physics and device
  application of two-dimensional transition metal dichalcogenide
  semiconductors},\ }\href {https://doi.org/10.1038/s41699-018-0074-2}
  {\bibfield  {journal} {\bibinfo  {journal} {npj 2D Mater Appl}\ }\textbf
  {\bibinfo {volume} {2}},\ \bibinfo {pages} {1} (\bibinfo {year} {2018})},\
  \bibinfo {note} {publisher: Nature Publishing Group}\BibitemShut {NoStop}%
\bibitem [{\citenamefont {Zollner}\ \emph {et~al.}(2019)\citenamefont
  {Zollner}, \citenamefont {Faria~Junior},\ and\ \citenamefont
  {Fabian}}]{Zollner2019}%
  \BibitemOpen
  \bibfield  {author} {\bibinfo {author} {\bibfnamefont {K.}~\bibnamefont
  {Zollner}}, \bibinfo {author} {\bibfnamefont {P.~E.}\ \bibnamefont
  {Faria~Junior}},\ and\ \bibinfo {author} {\bibfnamefont {J.}~\bibnamefont
  {Fabian}},\ }\bibfield  {title} {\bibinfo {title} {{Proximity exchange
  effects in ${\mathrm{MoSe}}_{2}$ and ${\mathrm{WSe}}_{2}$ heterostructures
  with ${\mathrm{CrI}}_{3}$: Twist angle, layer, and gate dependence}},\
  }\href@noop {} {\bibfield  {journal} {\bibinfo  {journal} {Phys. Rev. B}\
  }\textbf {\bibinfo {volume} {100}},\ \bibinfo {pages} {085128} (\bibinfo
  {year} {2019})}\BibitemShut {NoStop}%
\bibitem [{\citenamefont {Wang}(2013)}]{WangBook}%
  \BibitemOpen
  \bibfield  {author} {\bibinfo {author} {\bibfnamefont {Z.~M.}\ \bibnamefont
  {Wang}},\ }\href@noop {} {\emph {\bibinfo {title} {{MoS2}}}}\ (\bibinfo
  {publisher} {Springer Cham},\ \bibinfo {year} {2013})\BibitemShut {NoStop}%
\bibitem [{\citenamefont {Kolobov}\ and\ \citenamefont
  {Tominaga}(2016)}]{KolobovBook}%
  \BibitemOpen
  \bibfield  {author} {\bibinfo {author} {\bibfnamefont {A.~V.}\ \bibnamefont
  {Kolobov}}\ and\ \bibinfo {author} {\bibfnamefont {J.}~\bibnamefont
  {Tominaga}},\ }\href@noop {} {\emph {\bibinfo {title} {{Two-Dimensional
  Transition-Metal Dichalcogenides}}}}\ (\bibinfo  {publisher} {Springer},\
  \bibinfo {year} {2016})\BibitemShut {NoStop}%
\bibitem [{\citenamefont {Wee}\ \emph {et~al.}(2019)\citenamefont {Wee},
  \citenamefont {Chi},\ and\ \citenamefont {Goh}}]{WeeBook}%
  \BibitemOpen
  \bibfield  {author} {\bibinfo {author} {\bibfnamefont {A.~T.}\ \bibnamefont
  {Wee}}, \bibinfo {author} {\bibfnamefont {D.}~\bibnamefont {Chi}},\ and\
  \bibinfo {author} {\bibfnamefont {K.}~\bibnamefont {Goh}},\ }\href@noop {}
  {\emph {\bibinfo {title} {{2D Semiconductor Materials and Devices}}}}\
  (\bibinfo  {publisher} {Elsevier Science},\ \bibinfo {year}
  {2019})\BibitemShut {NoStop}%
\bibitem [{\citenamefont {Chernikov}\ \emph {et~al.}(2014)\citenamefont
  {Chernikov}, \citenamefont {Berkelbach}, \citenamefont {Hill}, \citenamefont
  {Rigosi}, \citenamefont {Li}, \citenamefont {Aslan}, \citenamefont
  {Reichman}, \citenamefont {Hybertsen},\ and\ \citenamefont
  {Heinz}}]{Chernikov2014}%
  \BibitemOpen
  \bibfield  {author} {\bibinfo {author} {\bibfnamefont {A.}~\bibnamefont
  {Chernikov}}, \bibinfo {author} {\bibfnamefont {T.~C.}\ \bibnamefont
  {Berkelbach}}, \bibinfo {author} {\bibfnamefont {H.~M.}\ \bibnamefont
  {Hill}}, \bibinfo {author} {\bibfnamefont {A.}~\bibnamefont {Rigosi}},
  \bibinfo {author} {\bibfnamefont {Y.}~\bibnamefont {Li}}, \bibinfo {author}
  {\bibfnamefont {B.}~\bibnamefont {Aslan}}, \bibinfo {author} {\bibfnamefont
  {D.~R.}\ \bibnamefont {Reichman}}, \bibinfo {author} {\bibfnamefont {M.~S.}\
  \bibnamefont {Hybertsen}},\ and\ \bibinfo {author} {\bibfnamefont {T.~F.}\
  \bibnamefont {Heinz}},\ }\bibfield  {title} {\bibinfo {title} {{Exciton
  Binding Energy and Nonhydrogenic Rydberg Series in Monolayer
  ${\mathrm{WS}}_{2}$}},\ }\href
  {https://doi.org/10.1103/PhysRevLett.113.076802} {\bibfield  {journal}
  {\bibinfo  {journal} {Phys. Rev. Lett.}\ }\textbf {\bibinfo {volume} {113}},\
  \bibinfo {pages} {076802} (\bibinfo {year} {2014})}\BibitemShut {NoStop}%
\bibitem [{\citenamefont {Back}\ \emph {et~al.}(2018)\citenamefont {Back},
  \citenamefont {Zeytinoglu}, \citenamefont {Ijaz}, \citenamefont {Kroner},\
  and\ \citenamefont {Imamo\ifmmode~\breve{g}\else \u{g}\fi{}lu}}]{Back2018}%
  \BibitemOpen
  \bibfield  {author} {\bibinfo {author} {\bibfnamefont {P.}~\bibnamefont
  {Back}}, \bibinfo {author} {\bibfnamefont {S.}~\bibnamefont {Zeytinoglu}},
  \bibinfo {author} {\bibfnamefont {A.}~\bibnamefont {Ijaz}}, \bibinfo {author}
  {\bibfnamefont {M.}~\bibnamefont {Kroner}},\ and\ \bibinfo {author}
  {\bibfnamefont {A.}~\bibnamefont {Imamo\ifmmode~\breve{g}\else
  \u{g}\fi{}lu}},\ }\bibfield  {title} {\bibinfo {title} {{Realization of an
  Electrically Tunable Narrow-Bandwidth Atomically Thin Mirror Using Monolayer
  ${\mathrm{MoSe}}_{2}$}},\ }\href
  {https://doi.org/10.1103/PhysRevLett.120.037401} {\bibfield  {journal}
  {\bibinfo  {journal} {Phys. Rev. Lett.}\ }\textbf {\bibinfo {volume} {120}},\
  \bibinfo {pages} {037401} (\bibinfo {year} {2018})}\BibitemShut {NoStop}%
\bibitem [{\citenamefont {Mak}\ \emph {et~al.}(2014)\citenamefont {Mak},
  \citenamefont {McGill}, \citenamefont {Park},\ and\ \citenamefont
  {McEuen}}]{Mak2014}%
  \BibitemOpen
  \bibfield  {author} {\bibinfo {author} {\bibfnamefont {K.~F.}\ \bibnamefont
  {Mak}}, \bibinfo {author} {\bibfnamefont {K.~L.}\ \bibnamefont {McGill}},
  \bibinfo {author} {\bibfnamefont {J.}~\bibnamefont {Park}},\ and\ \bibinfo
  {author} {\bibfnamefont {P.~L.}\ \bibnamefont {McEuen}},\ }\bibfield  {title}
  {\bibinfo {title} {{The valley Hall effect in MoS$_{2}$ transistors}},\
  }\href {https://doi.org/10.1126/science.1250140} {\bibfield  {journal}
  {\bibinfo  {journal} {Science}\ }\textbf {\bibinfo {volume} {344}},\ \bibinfo
  {pages} {1489} (\bibinfo {year} {2014})}\BibitemShut {NoStop}%
\bibitem [{\citenamefont {Habe}\ and\ \citenamefont
  {Koshino}(2017)}]{Habe2017}%
  \BibitemOpen
  \bibfield  {author} {\bibinfo {author} {\bibfnamefont {T.}~\bibnamefont
  {Habe}}\ and\ \bibinfo {author} {\bibfnamefont {M.}~\bibnamefont {Koshino}},\
  }\bibfield  {title} {\bibinfo {title} {{Anomalous Hall effect in $2H$-phase
  $M{X}_{2}$ transition-metal dichalcogenide monolayers on ferromagnetic
  substrates ($M$ = Mo, W, and $X$ = S, Se, Te)}},\ }\href
  {https://doi.org/10.1103/PhysRevB.96.085411} {\bibfield  {journal} {\bibinfo
  {journal} {Phys. Rev. B}\ }\textbf {\bibinfo {volume} {96}},\ \bibinfo
  {pages} {085411} (\bibinfo {year} {2017})}\BibitemShut {NoStop}%
\bibitem [{\citenamefont {Schulman}\ \emph {et~al.}(2018)\citenamefont
  {Schulman}, \citenamefont {Arnold},\ and\ \citenamefont
  {Das}}]{schulman2018}%
  \BibitemOpen
  \bibfield  {author} {\bibinfo {author} {\bibfnamefont {D.~S.}\ \bibnamefont
  {Schulman}}, \bibinfo {author} {\bibfnamefont {A.~J.}\ \bibnamefont
  {Arnold}},\ and\ \bibinfo {author} {\bibfnamefont {S.}~\bibnamefont {Das}},\
  }\bibfield  {title} {\bibinfo {title} {{Contact engineering for 2D materials
  and devices}},\ }\href@noop {} {\bibfield  {journal} {\bibinfo  {journal}
  {Chemical Society Reviews}\ }\textbf {\bibinfo {volume} {47}},\ \bibinfo
  {pages} {3037} (\bibinfo {year} {2018})}\BibitemShut {NoStop}%
\bibitem [{\citenamefont {Allain}\ \emph {et~al.}(2015)\citenamefont {Allain},
  \citenamefont {Kang}, \citenamefont {Banerjee},\ and\ \citenamefont
  {Kis}}]{allain2015}%
  \BibitemOpen
  \bibfield  {author} {\bibinfo {author} {\bibfnamefont {A.}~\bibnamefont
  {Allain}}, \bibinfo {author} {\bibfnamefont {J.}~\bibnamefont {Kang}},
  \bibinfo {author} {\bibfnamefont {K.}~\bibnamefont {Banerjee}},\ and\
  \bibinfo {author} {\bibfnamefont {A.}~\bibnamefont {Kis}},\ }\bibfield
  {title} {\bibinfo {title} {Electrical contacts to two-dimensional
  semiconductors},\ }\href@noop {} {\bibfield  {journal} {\bibinfo  {journal}
  {Nature Materials}\ }\textbf {\bibinfo {volume} {14}},\ \bibinfo {pages}
  {1195} (\bibinfo {year} {2015})}\BibitemShut {NoStop}%
\bibitem [{\citenamefont {Zheng}\ \emph {et~al.}(2021)\citenamefont {Zheng},
  \citenamefont {Gao}, \citenamefont {Han},\ and\ \citenamefont
  {Chen}}]{zheng2021}%
  \BibitemOpen
  \bibfield  {author} {\bibinfo {author} {\bibfnamefont {Y.}~\bibnamefont
  {Zheng}}, \bibinfo {author} {\bibfnamefont {J.}~\bibnamefont {Gao}}, \bibinfo
  {author} {\bibfnamefont {C.}~\bibnamefont {Han}},\ and\ \bibinfo {author}
  {\bibfnamefont {W.}~\bibnamefont {Chen}},\ }\bibfield  {title} {\bibinfo
  {title} {{Ohmic Contact Engineering for Two-Dimensional Materials}},\
  }\href@noop {} {\bibfield  {journal} {\bibinfo  {journal} {Cell Reports
  Physical Science}\ }\textbf {\bibinfo {volume} {2}},\ \bibinfo {pages}
  {100298} (\bibinfo {year} {2021})}\BibitemShut {NoStop}%
\bibitem [{\citenamefont {Shen}\ \emph {et~al.}(2021)\citenamefont {Shen},
  \citenamefont {Su}, \citenamefont {Lin}, \citenamefont {Chou}, \citenamefont
  {Cheng}, \citenamefont {Park}, \citenamefont {Chiu}, \citenamefont {Lu},
  \citenamefont {Tang}, \citenamefont {Tavakoli}, \citenamefont {Pitner},
  \citenamefont {Ji}, \citenamefont {Cai}, \citenamefont {Mao}, \citenamefont
  {Wang}, \citenamefont {Tung}, \citenamefont {Li}, \citenamefont {Bokor},
  \citenamefont {Zettl}, \citenamefont {Wu}, \citenamefont {Palacios},
  \citenamefont {Li},\ and\ \citenamefont {Kong}}]{shen2021}%
  \BibitemOpen
  \bibfield  {author} {\bibinfo {author} {\bibfnamefont {P.-C.}\ \bibnamefont
  {Shen}}, \bibinfo {author} {\bibfnamefont {C.}~\bibnamefont {Su}}, \bibinfo
  {author} {\bibfnamefont {Y.}~\bibnamefont {Lin}}, \bibinfo {author}
  {\bibfnamefont {A.-S.}\ \bibnamefont {Chou}}, \bibinfo {author}
  {\bibfnamefont {C.-C.}\ \bibnamefont {Cheng}}, \bibinfo {author}
  {\bibfnamefont {J.-H.}\ \bibnamefont {Park}}, \bibinfo {author}
  {\bibfnamefont {M.-H.}\ \bibnamefont {Chiu}}, \bibinfo {author}
  {\bibfnamefont {A.-Y.}\ \bibnamefont {Lu}}, \bibinfo {author} {\bibfnamefont
  {H.-L.}\ \bibnamefont {Tang}}, \bibinfo {author} {\bibfnamefont {M.~M.}\
  \bibnamefont {Tavakoli}}, \bibinfo {author} {\bibfnamefont {G.}~\bibnamefont
  {Pitner}}, \bibinfo {author} {\bibfnamefont {X.}~\bibnamefont {Ji}}, \bibinfo
  {author} {\bibfnamefont {Z.}~\bibnamefont {Cai}}, \bibinfo {author}
  {\bibfnamefont {N.}~\bibnamefont {Mao}}, \bibinfo {author} {\bibfnamefont
  {J.}~\bibnamefont {Wang}}, \bibinfo {author} {\bibfnamefont {V.}~\bibnamefont
  {Tung}}, \bibinfo {author} {\bibfnamefont {J.}~\bibnamefont {Li}}, \bibinfo
  {author} {\bibfnamefont {J.}~\bibnamefont {Bokor}}, \bibinfo {author}
  {\bibfnamefont {A.}~\bibnamefont {Zettl}}, \bibinfo {author} {\bibfnamefont
  {C.-I.}\ \bibnamefont {Wu}}, \bibinfo {author} {\bibfnamefont
  {T.}~\bibnamefont {Palacios}}, \bibinfo {author} {\bibfnamefont {L.-J.}\
  \bibnamefont {Li}},\ and\ \bibinfo {author} {\bibfnamefont {J.}~\bibnamefont
  {Kong}},\ }\bibfield  {title} {\bibinfo {title} {Ultralow contact resistance
  between semimetal and monolayer semiconductors},\ }\href@noop {} {\bibfield
  {journal} {\bibinfo  {journal} {Nature}\ }\textbf {\bibinfo {volume} {593}},\
  \bibinfo {pages} {211} (\bibinfo {year} {2021})}\BibitemShut {NoStop}%
\bibitem [{\citenamefont {Li}\ \emph {et~al.}(2023)\citenamefont {Li},
  \citenamefont {Gong}, \citenamefont {Yu}, \citenamefont {Ma}, \citenamefont
  {Sun}, \citenamefont {Gao}, \citenamefont {K\"{o}ro\u{g}lu}, \citenamefont
  {Wang}, \citenamefont {Liu}, \citenamefont {Li}, \citenamefont {Ning},
  \citenamefont {Fan}, \citenamefont {Xu}, \citenamefont {Tu}, \citenamefont
  {Xu}, \citenamefont {Sun}, \citenamefont {Wang}, \citenamefont {Lu},
  \citenamefont {Ni}, \citenamefont {Li}, \citenamefont {Duan}, \citenamefont
  {Wang}, \citenamefont {Nie}, \citenamefont {Qiu}, \citenamefont {Shi},
  \citenamefont {Pop}, \citenamefont {Wang},\ and\ \citenamefont
  {Wang}}]{li_approaching_2023}%
  \BibitemOpen
  \bibfield  {author} {\bibinfo {author} {\bibfnamefont {W.}~\bibnamefont
  {Li}}, \bibinfo {author} {\bibfnamefont {X.}~\bibnamefont {Gong}}, \bibinfo
  {author} {\bibfnamefont {Z.}~\bibnamefont {Yu}}, \bibinfo {author}
  {\bibfnamefont {L.}~\bibnamefont {Ma}}, \bibinfo {author} {\bibfnamefont
  {W.}~\bibnamefont {Sun}}, \bibinfo {author} {\bibfnamefont {S.}~\bibnamefont
  {Gao}}, \bibinfo {author} {\bibfnamefont {C.}~\bibnamefont
  {K\"{o}ro\u{g}lu}}, \bibinfo {author} {\bibfnamefont {W.}~\bibnamefont
  {Wang}}, \bibinfo {author} {\bibfnamefont {L.}~\bibnamefont {Liu}}, \bibinfo
  {author} {\bibfnamefont {T.}~\bibnamefont {Li}}, \bibinfo {author}
  {\bibfnamefont {H.}~\bibnamefont {Ning}}, \bibinfo {author} {\bibfnamefont
  {D.}~\bibnamefont {Fan}}, \bibinfo {author} {\bibfnamefont {Y.}~\bibnamefont
  {Xu}}, \bibinfo {author} {\bibfnamefont {X.}~\bibnamefont {Tu}}, \bibinfo
  {author} {\bibfnamefont {T.}~\bibnamefont {Xu}}, \bibinfo {author}
  {\bibfnamefont {L.}~\bibnamefont {Sun}}, \bibinfo {author} {\bibfnamefont
  {W.}~\bibnamefont {Wang}}, \bibinfo {author} {\bibfnamefont {J.}~\bibnamefont
  {Lu}}, \bibinfo {author} {\bibfnamefont {Z.}~\bibnamefont {Ni}}, \bibinfo
  {author} {\bibfnamefont {J.}~\bibnamefont {Li}}, \bibinfo {author}
  {\bibfnamefont {X.}~\bibnamefont {Duan}}, \bibinfo {author} {\bibfnamefont
  {P.}~\bibnamefont {Wang}}, \bibinfo {author} {\bibfnamefont {Y.}~\bibnamefont
  {Nie}}, \bibinfo {author} {\bibfnamefont {H.}~\bibnamefont {Qiu}}, \bibinfo
  {author} {\bibfnamefont {Y.}~\bibnamefont {Shi}}, \bibinfo {author}
  {\bibfnamefont {E.}~\bibnamefont {Pop}}, \bibinfo {author} {\bibfnamefont
  {J.}~\bibnamefont {Wang}},\ and\ \bibinfo {author} {\bibfnamefont
  {X.}~\bibnamefont {Wang}},\ }\bibfield  {title} {\bibinfo {title}
  {Approaching the quantum limit in two-dimensional semiconductor contacts},\
  }\href {https://doi.org/10.1038/s41586-022-05431-4} {\bibfield  {journal}
  {\bibinfo  {journal} {Nature}\ }\textbf {\bibinfo {volume} {613}},\ \bibinfo
  {pages} {274} (\bibinfo {year} {2023})},\ \bibinfo {note} {publisher: Nature
  Publishing Group}\BibitemShut {NoStop}%
\bibitem [{\citenamefont {Lee}\ \emph {et~al.}(2024)\citenamefont {Lee},
  \citenamefont {Wang}, \citenamefont {Shin}, \citenamefont {Ali},
  \citenamefont {Ngo}, \citenamefont {Hwang}, \citenamefont {Kim},
  \citenamefont {Yeom}, \citenamefont {Watanabe}, \citenamefont {Taniguchi},\
  and\ \citenamefont {Yoo}}]{Lee2024}%
  \BibitemOpen
  \bibfield  {author} {\bibinfo {author} {\bibfnamefont {S.}~\bibnamefont
  {Lee}}, \bibinfo {author} {\bibfnamefont {X.}~\bibnamefont {Wang}}, \bibinfo
  {author} {\bibfnamefont {H.}~\bibnamefont {Shin}}, \bibinfo {author}
  {\bibfnamefont {N.}~\bibnamefont {Ali}}, \bibinfo {author} {\bibfnamefont
  {T.~D.}\ \bibnamefont {Ngo}}, \bibinfo {author} {\bibfnamefont
  {E.}~\bibnamefont {Hwang}}, \bibinfo {author} {\bibfnamefont {G.-H.}\
  \bibnamefont {Kim}}, \bibinfo {author} {\bibfnamefont {G.~Y.}\ \bibnamefont
  {Yeom}}, \bibinfo {author} {\bibfnamefont {K.}~\bibnamefont {Watanabe}},
  \bibinfo {author} {\bibfnamefont {T.}~\bibnamefont {Taniguchi}},\ and\
  \bibinfo {author} {\bibfnamefont {W.~J.}\ \bibnamefont {Yoo}},\ }\bibfield
  {title} {\bibinfo {title} {{Semi-Metal Edge Contact for Barrier-Free Carrier
  Transport in MoS2 Field Effect Transistors}},\ }\href
  {https://doi.org/10.1021/acsaelm.4c00250} {\bibfield  {journal} {\bibinfo
  {journal} {ACS Applied Electronic Materials}\ }\textbf {\bibinfo {volume}
  {6}},\ \bibinfo {pages} {4149} (\bibinfo {year} {2024})}\BibitemShut
  {NoStop}%
\bibitem [{\citenamefont {Su}\ \emph {et~al.}(2023)\citenamefont {Su},
  \citenamefont {Li}, \citenamefont {Wang}, \citenamefont {Zhao}, \citenamefont
  {Cao},\ and\ \citenamefont {Ang}}]{Su2023}%
  \BibitemOpen
  \bibfield  {author} {\bibinfo {author} {\bibfnamefont {T.}~\bibnamefont
  {Su}}, \bibinfo {author} {\bibfnamefont {Y.}~\bibnamefont {Li}}, \bibinfo
  {author} {\bibfnamefont {Q.}~\bibnamefont {Wang}}, \bibinfo {author}
  {\bibfnamefont {W.}~\bibnamefont {Zhao}}, \bibinfo {author} {\bibfnamefont
  {L.}~\bibnamefont {Cao}},\ and\ \bibinfo {author} {\bibfnamefont {Y.~S.}\
  \bibnamefont {Ang}},\ }\bibfield  {title} {\bibinfo {title} {{Semimetal
  contacts to monolayer semiconductor: weak metalization as an effective
  mechanism to Schottky barrier lowering}},\ }\href@noop {} {\bibfield
  {journal} {\bibinfo  {journal} {Journal of Physics D: Applied Physics}\
  }\textbf {\bibinfo {volume} {56}} (\bibinfo {year} {2023})}\BibitemShut
  {NoStop}%
\bibitem [{\citenamefont {Turunen}\ \emph {et~al.}(2023)\citenamefont
  {Turunen}, \citenamefont {Fernandez}, \citenamefont {Akkanen}, \citenamefont
  {Seppänen},\ and\ \citenamefont {Sun}}]{Turunen2023}%
  \BibitemOpen
  \bibfield  {author} {\bibinfo {author} {\bibfnamefont {M.}~\bibnamefont
  {Turunen}}, \bibinfo {author} {\bibfnamefont {H.}~\bibnamefont {Fernandez}},
  \bibinfo {author} {\bibfnamefont {S.-T.}\ \bibnamefont {Akkanen}}, \bibinfo
  {author} {\bibfnamefont {H.}~\bibnamefont {Seppänen}},\ and\ \bibinfo
  {author} {\bibfnamefont {Z.}~\bibnamefont {Sun}},\ }\bibfield  {title}
  {\bibinfo {title} {Effects of atomic layer deposition on the optical
  properties of two-dimensional transition metal dichalcogenide monolayers},\
  }\href {https://doi.org/10.1088/2053-1583/acf1ad} {\bibfield  {journal}
  {\bibinfo  {journal} {2D Materials}\ }\textbf {\bibinfo {volume} {10}},\
  \bibinfo {pages} {045018} (\bibinfo {year} {2023})}\BibitemShut {NoStop}%
\bibitem [{\citenamefont {Fan}\ \emph {et~al.}(2018)\citenamefont {Fan},
  \citenamefont {Guo}, \citenamefont {Zhu}, \citenamefont {Xu}, \citenamefont
  {Wu}, \citenamefont {Yuan},\ and\ \citenamefont {Qin}}]{Fan2018}%
  \BibitemOpen
  \bibfield  {author} {\bibinfo {author} {\bibfnamefont {Y.}~\bibnamefont
  {Fan}}, \bibinfo {author} {\bibfnamefont {C.}~\bibnamefont {Guo}}, \bibinfo
  {author} {\bibfnamefont {Z.}~\bibnamefont {Zhu}}, \bibinfo {author}
  {\bibfnamefont {W.}~\bibnamefont {Xu}}, \bibinfo {author} {\bibfnamefont
  {F.}~\bibnamefont {Wu}}, \bibinfo {author} {\bibfnamefont {X.}~\bibnamefont
  {Yuan}},\ and\ \bibinfo {author} {\bibfnamefont {S.}~\bibnamefont {Qin}},\
  }\bibfield  {title} {\bibinfo {title} {{Monolayer-graphene-based broadband
  and wide-angle perfect absorption structures in the near infrared}},\
  }\href@noop {} {\bibfield  {journal} {\bibinfo  {journal} {Scientific
  Reports}\ }\textbf {\bibinfo {volume} {8}} (\bibinfo {year}
  {2018})}\BibitemShut {NoStop}%
\bibitem [{\citenamefont {Johnson}\ and\ \citenamefont
  {Christy}(1972)}]{Johnson1972}%
  \BibitemOpen
  \bibfield  {author} {\bibinfo {author} {\bibfnamefont {P.~B.}\ \bibnamefont
  {Johnson}}\ and\ \bibinfo {author} {\bibfnamefont {R.~W.}\ \bibnamefont
  {Christy}},\ }\bibfield  {title} {\bibinfo {title} {{Optical Constants of the
  Noble Metals}},\ }\href {https://doi.org/10.1103/PhysRevB.6.4370} {\bibfield
  {journal} {\bibinfo  {journal} {Phys. Rev. B}\ }\textbf {\bibinfo {volume}
  {6}},\ \bibinfo {pages} {4370} (\bibinfo {year} {1972})}\BibitemShut
  {NoStop}%
\bibitem [{\citenamefont {Ahmed}\ \emph {et~al.}(2020)\citenamefont {Ahmed},
  \citenamefont {Roy}, \citenamefont {Kakkar}, \citenamefont {Pradhan},\ and\
  \citenamefont {Ghosh}}]{Ahmed2020}%
  \BibitemOpen
  \bibfield  {author} {\bibinfo {author} {\bibfnamefont {T.}~\bibnamefont
  {Ahmed}}, \bibinfo {author} {\bibfnamefont {K.}~\bibnamefont {Roy}}, \bibinfo
  {author} {\bibfnamefont {S.}~\bibnamefont {Kakkar}}, \bibinfo {author}
  {\bibfnamefont {A.}~\bibnamefont {Pradhan}},\ and\ \bibinfo {author}
  {\bibfnamefont {A.}~\bibnamefont {Ghosh}},\ }\bibfield  {title} {\bibinfo
  {title} {{Interplay of charge transfer and disorder in optoelectronic
  response in Graphene/hBN/MoS2 van der Waals heterostructures}},\ }\href
  {https://doi.org/10.1088/2053-1583/ab771f} {\bibfield  {journal} {\bibinfo
  {journal} {2D Materials}\ }\textbf {\bibinfo {volume} {7}},\ \bibinfo {pages}
  {025043} (\bibinfo {year} {2020})}\BibitemShut {NoStop}%
\bibitem [{\citenamefont {Grzeszczyk}\ \emph {et~al.}(2020)\citenamefont
  {Grzeszczyk}, \citenamefont {Molas}, \citenamefont {Nogajewski},
  \citenamefont {Bartoš}, \citenamefont {Bogucki}, \citenamefont {Faugeras},
  \citenamefont {Kossacki}, \citenamefont {Babiński},\ and\ \citenamefont
  {Potemski}}]{Grzeszczyk2020}%
  \BibitemOpen
  \bibfield  {author} {\bibinfo {author} {\bibfnamefont {M.}~\bibnamefont
  {Grzeszczyk}}, \bibinfo {author} {\bibfnamefont {M.~R.}\ \bibnamefont
  {Molas}}, \bibinfo {author} {\bibfnamefont {K.}~\bibnamefont {Nogajewski}},
  \bibinfo {author} {\bibfnamefont {M.}~\bibnamefont {Bartoš}}, \bibinfo
  {author} {\bibfnamefont {A.}~\bibnamefont {Bogucki}}, \bibinfo {author}
  {\bibfnamefont {C.}~\bibnamefont {Faugeras}}, \bibinfo {author}
  {\bibfnamefont {P.}~\bibnamefont {Kossacki}}, \bibinfo {author}
  {\bibfnamefont {A.}~\bibnamefont {Babiński}},\ and\ \bibinfo {author}
  {\bibfnamefont {M.}~\bibnamefont {Potemski}},\ }\bibfield  {title} {\bibinfo
  {title} {{The effect of metallic substrates on the optical properties of
  monolayer MoSe$_{2}$}},\ }\href@noop {} {\bibfield  {journal} {\bibinfo
  {journal} {Scientific Reports}\ }\textbf {\bibinfo {volume} {10}} (\bibinfo
  {year} {2020})}\BibitemShut {NoStop}%
\bibitem [{\citenamefont {Jadczak}\ \emph {et~al.}(2017)\citenamefont
  {Jadczak}, \citenamefont {Kutrowska-Girzycka}, \citenamefont {Kapuściński},
  \citenamefont {Huang}, \citenamefont {Wójs},\ and\ \citenamefont
  {Bryja}}]{Jadczak2017}%
  \BibitemOpen
  \bibfield  {author} {\bibinfo {author} {\bibfnamefont {J.}~\bibnamefont
  {Jadczak}}, \bibinfo {author} {\bibfnamefont {J.}~\bibnamefont
  {Kutrowska-Girzycka}}, \bibinfo {author} {\bibfnamefont {P.}~\bibnamefont
  {Kapuściński}}, \bibinfo {author} {\bibfnamefont {Y.~S.}\ \bibnamefont
  {Huang}}, \bibinfo {author} {\bibfnamefont {A.}~\bibnamefont {Wójs}},\ and\
  \bibinfo {author} {\bibfnamefont {L.}~\bibnamefont {Bryja}},\ }\bibfield
  {title} {\bibinfo {title} {{Probing of free and localized excitons and trions
  in atomically thin WSe2, WS2, MoSe2 and MoS2 in photoluminescence and
  reflectivity experiments}},\ }\href@noop {} {\bibfield  {journal} {\bibinfo
  {journal} {Nanotechnology}\ }\textbf {\bibinfo {volume} {28}},\ \bibinfo
  {pages} {395702} (\bibinfo {year} {2017})}\BibitemShut {NoStop}%
\bibitem [{\citenamefont {Munson}\ \emph {et~al.}(2024)\citenamefont {Munson},
  \citenamefont {Torsi}, \citenamefont {Mathela}, \citenamefont {Feidler},
  \citenamefont {Lin}, \citenamefont {Robinson},\ and\ \citenamefont
  {Asbury}}]{Munson2024}%
  \BibitemOpen
  \bibfield  {author} {\bibinfo {author} {\bibfnamefont {K.~T.}\ \bibnamefont
  {Munson}}, \bibinfo {author} {\bibfnamefont {R.}~\bibnamefont {Torsi}},
  \bibinfo {author} {\bibfnamefont {S.}~\bibnamefont {Mathela}}, \bibinfo
  {author} {\bibfnamefont {M.~A.}\ \bibnamefont {Feidler}}, \bibinfo {author}
  {\bibfnamefont {Y.-C.}\ \bibnamefont {Lin}}, \bibinfo {author} {\bibfnamefont
  {J.~A.}\ \bibnamefont {Robinson}},\ and\ \bibinfo {author} {\bibfnamefont
  {J.~B.}\ \bibnamefont {Asbury}},\ }\bibfield  {title} {\bibinfo {title}
  {{Influence of Substrate-Induced Charge Doping on Defect-Related Excitonic
  Emission in Monolayer MoS2}},\ }\href
  {https://doi.org/10.1021/acs.jpclett.4c01578} {\bibfield  {journal} {\bibinfo
   {journal} {The Journal of Physical Chemistry Letters}\ }\textbf {\bibinfo
  {volume} {15}},\ \bibinfo {pages} {7850} (\bibinfo {year}
  {2024})}\BibitemShut {NoStop}%
\bibitem [{\citenamefont {Park}\ \emph {et~al.}(2023)\citenamefont {Park},
  \citenamefont {Li}, \citenamefont {Jung}, \citenamefont {Singh},
  \citenamefont {Baik}, \citenamefont {Lee}, \citenamefont {Oh}, \citenamefont
  {Kim}, \citenamefont {Lee}, \citenamefont {Woo}, \citenamefont {Park},
  \citenamefont {Kim}, \citenamefont {Lee}, \citenamefont {Lee},\ and\
  \citenamefont {Hwang}}]{Park2023}%
  \BibitemOpen
  \bibfield  {author} {\bibinfo {author} {\bibfnamefont {Y.}~\bibnamefont
  {Park}}, \bibinfo {author} {\bibfnamefont {N.}~\bibnamefont {Li}}, \bibinfo
  {author} {\bibfnamefont {D.}~\bibnamefont {Jung}}, \bibinfo {author}
  {\bibfnamefont {L.~T.}\ \bibnamefont {Singh}}, \bibinfo {author}
  {\bibfnamefont {J.}~\bibnamefont {Baik}}, \bibinfo {author} {\bibfnamefont
  {E.}~\bibnamefont {Lee}}, \bibinfo {author} {\bibfnamefont {D.}~\bibnamefont
  {Oh}}, \bibinfo {author} {\bibfnamefont {Y.~D.}\ \bibnamefont {Kim}},
  \bibinfo {author} {\bibfnamefont {J.~Y.}\ \bibnamefont {Lee}}, \bibinfo
  {author} {\bibfnamefont {J.}~\bibnamefont {Woo}}, \bibinfo {author}
  {\bibfnamefont {S.}~\bibnamefont {Park}}, \bibinfo {author} {\bibfnamefont
  {H.}~\bibnamefont {Kim}}, \bibinfo {author} {\bibfnamefont {G.}~\bibnamefont
  {Lee}}, \bibinfo {author} {\bibfnamefont {G.}~\bibnamefont {Lee}},\ and\
  \bibinfo {author} {\bibfnamefont {C.-C.}\ \bibnamefont {Hwang}},\ }\bibfield
  {title} {\bibinfo {title} {{Unveiling the origin of n-type doping of natural
  MoS2: carbon}},\ }\href@noop {} {\bibfield  {journal} {\bibinfo  {journal}
  {npj 2D Materials and Applications}\ }\textbf {\bibinfo {volume} {7}}
  (\bibinfo {year} {2023})}\BibitemShut {NoStop}%
\bibitem [{\citenamefont {Courtade}\ \emph {et~al.}(2017)\citenamefont
  {Courtade}, \citenamefont {Semina}, \citenamefont {Manca}, \citenamefont
  {Glazov}, \citenamefont {Robert}, \citenamefont {Cadiz}, \citenamefont
  {Wang}, \citenamefont {Taniguchi}, \citenamefont {Watanabe}, \citenamefont
  {Pierre}, \citenamefont {Escoffier}, \citenamefont {Ivchenko}, \citenamefont
  {Renucci}, \citenamefont {Marie}, \citenamefont {Amand},\ and\ \citenamefont
  {Urbaszek}}]{Courtade2017}%
  \BibitemOpen
  \bibfield  {author} {\bibinfo {author} {\bibfnamefont {E.}~\bibnamefont
  {Courtade}}, \bibinfo {author} {\bibfnamefont {M.}~\bibnamefont {Semina}},
  \bibinfo {author} {\bibfnamefont {M.}~\bibnamefont {Manca}}, \bibinfo
  {author} {\bibfnamefont {M.~M.}\ \bibnamefont {Glazov}}, \bibinfo {author}
  {\bibfnamefont {C.}~\bibnamefont {Robert}}, \bibinfo {author} {\bibfnamefont
  {F.}~\bibnamefont {Cadiz}}, \bibinfo {author} {\bibfnamefont
  {G.}~\bibnamefont {Wang}}, \bibinfo {author} {\bibfnamefont {T.}~\bibnamefont
  {Taniguchi}}, \bibinfo {author} {\bibfnamefont {K.}~\bibnamefont {Watanabe}},
  \bibinfo {author} {\bibfnamefont {M.}~\bibnamefont {Pierre}}, \bibinfo
  {author} {\bibfnamefont {W.}~\bibnamefont {Escoffier}}, \bibinfo {author}
  {\bibfnamefont {E.~L.}\ \bibnamefont {Ivchenko}}, \bibinfo {author}
  {\bibfnamefont {P.}~\bibnamefont {Renucci}}, \bibinfo {author} {\bibfnamefont
  {X.}~\bibnamefont {Marie}}, \bibinfo {author} {\bibfnamefont
  {T.}~\bibnamefont {Amand}},\ and\ \bibinfo {author} {\bibfnamefont
  {B.}~\bibnamefont {Urbaszek}},\ }\bibfield  {title} {\bibinfo {title}
  {{Charged excitons in monolayer ${\mathrm{WSe}}_{2}$: Experiment and
  theory}},\ }\href {https://doi.org/10.1103/PhysRevB.96.085302} {\bibfield
  {journal} {\bibinfo  {journal} {Phys. Rev. B}\ }\textbf {\bibinfo {volume}
  {96}},\ \bibinfo {pages} {085302} (\bibinfo {year} {2017})}\BibitemShut
  {NoStop}%
\bibitem [{\citenamefont {Sidler}\ \emph {et~al.}(2017)\citenamefont {Sidler},
  \citenamefont {Back}, \citenamefont {Cotlet}, \citenamefont {Srivastava},
  \citenamefont {Fink}, \citenamefont {Kroner}, \citenamefont {Demler},\ and\
  \citenamefont {Imamoglu}}]{Sidler2017}%
  \BibitemOpen
  \bibfield  {author} {\bibinfo {author} {\bibfnamefont {M.}~\bibnamefont
  {Sidler}}, \bibinfo {author} {\bibfnamefont {P.}~\bibnamefont {Back}},
  \bibinfo {author} {\bibfnamefont {O.}~\bibnamefont {Cotlet}}, \bibinfo
  {author} {\bibfnamefont {A.}~\bibnamefont {Srivastava}}, \bibinfo {author}
  {\bibfnamefont {T.}~\bibnamefont {Fink}}, \bibinfo {author} {\bibfnamefont
  {M.}~\bibnamefont {Kroner}}, \bibinfo {author} {\bibfnamefont
  {E.}~\bibnamefont {Demler}},\ and\ \bibinfo {author} {\bibfnamefont
  {A.}~\bibnamefont {Imamoglu}},\ }\bibfield  {title} {\bibinfo {title} {Fermi
  polaron-polaritons in charge-tunable atomically thin semiconductors},\
  }\href@noop {} {\bibfield  {journal} {\bibinfo  {journal} {Nature Physics}\
  }\textbf {\bibinfo {volume} {13}},\ \bibinfo {pages} {255} (\bibinfo {year}
  {2017})}\BibitemShut {NoStop}%
\bibitem [{\citenamefont {Tran}\ \emph {et~al.}(2016)\citenamefont {Tran},
  \citenamefont {Kim},\ and\ \citenamefont {Lee}}]{tran2016}%
  \BibitemOpen
  \bibfield  {author} {\bibinfo {author} {\bibfnamefont {M.~D.}\ \bibnamefont
  {Tran}}, \bibinfo {author} {\bibfnamefont {J.-H.}\ \bibnamefont {Kim}},\ and\
  \bibinfo {author} {\bibfnamefont {Y.~H.}\ \bibnamefont {Lee}},\ }\bibfield
  {title} {\bibinfo {title} {Tailoring photoluminescence of monolayer
  transition metal dichalcogenides},\ }\href@noop {} {\bibfield  {journal}
  {\bibinfo  {journal} {Current Applied Physics}\ }\textbf {\bibinfo {volume}
  {16}},\ \bibinfo {pages} {1159} (\bibinfo {year} {2016})}\BibitemShut
  {NoStop}%
\bibitem [{\citenamefont {Zielińska}\ \emph {et~al.}(2025)\citenamefont
  {Zielińska}, \citenamefont {Prażmowska-Czajka}, \citenamefont {Dyksik},
  \citenamefont {Eroms}, \citenamefont {Weiss}, \citenamefont {Paszkiewicz},\
  and\ \citenamefont {Ciorga}}]{Zielinska2025}%
  \BibitemOpen
  \bibfield  {author} {\bibinfo {author} {\bibfnamefont {A.}~\bibnamefont
  {Zielińska}}, \bibinfo {author} {\bibfnamefont {J.}~\bibnamefont
  {Prażmowska-Czajka}}, \bibinfo {author} {\bibfnamefont {M.}~\bibnamefont
  {Dyksik}}, \bibinfo {author} {\bibfnamefont {J.}~\bibnamefont {Eroms}},
  \bibinfo {author} {\bibfnamefont {D.}~\bibnamefont {Weiss}}, \bibinfo
  {author} {\bibfnamefont {R.}~\bibnamefont {Paszkiewicz}},\ and\ \bibinfo
  {author} {\bibfnamefont {M.}~\bibnamefont {Ciorga}},\ }\bibfield  {title}
  {\bibinfo {title} {{Study of gold and bismuth electrical contacts to a MoS2
  monolayer}},\ }\href
  {https://doi.org/https://doi.org/10.1016/j.ssc.2024.115824} {\bibfield
  {journal} {\bibinfo  {journal} {Solid State Communications}\ }\textbf
  {\bibinfo {volume} {397}},\ \bibinfo {pages} {115824} (\bibinfo {year}
  {2025})}\BibitemShut {NoStop}%
\bibitem [{\citenamefont {Feng}\ \emph {et~al.}(2024)\citenamefont {Feng},
  \citenamefont {Yu}, \citenamefont {Guo}, \citenamefont {Li}, \citenamefont
  {Zhang},\ and\ \citenamefont {Ang}}]{feng_direct_2024}%
  \BibitemOpen
  \bibfield  {author} {\bibinfo {author} {\bibfnamefont {X.}~\bibnamefont
  {Feng}}, \bibinfo {author} {\bibfnamefont {Z.~G.}\ \bibnamefont {Yu}},
  \bibinfo {author} {\bibfnamefont {H.}~\bibnamefont {Guo}}, \bibinfo {author}
  {\bibfnamefont {Y.}~\bibnamefont {Li}}, \bibinfo {author} {\bibfnamefont
  {Y.-W.}\ \bibnamefont {Zhang}},\ and\ \bibinfo {author} {\bibfnamefont
  {K.-W.}\ \bibnamefont {Ang}},\ }\bibfield  {title} {\bibinfo {title} {Direct
  {Observation} of {Semimetal} {Contact} {Induced} {Charge} {Doping} and
  {Strain} {Effect} in {CVD}-{Grown} {Monolayer} {MoS2} {Transistors}},\ }\href
  {https://doi.org/10.1002/aelm.202300820} {\bibfield  {journal} {\bibinfo
  {journal} {Advanced Electronic Materials}\ }\textbf {\bibinfo {volume}
  {10}},\ \bibinfo {pages} {2300820} (\bibinfo {year} {2024})},\ \bibinfo
  {note} {\_eprint:
  https://onlinelibrary.wiley.com/doi/pdf/10.1002/aelm.202300820}\BibitemShut
  {NoStop}%
\bibitem [{\citenamefont {Sovizi}\ and\ \citenamefont
  {Szoszkiewicz}(2022)}]{sovizi_single_2022}%
  \BibitemOpen
  \bibfield  {author} {\bibinfo {author} {\bibfnamefont {S.}~\bibnamefont
  {Sovizi}}\ and\ \bibinfo {author} {\bibfnamefont {R.}~\bibnamefont
  {Szoszkiewicz}},\ }\bibfield  {title} {\bibinfo {title} {Single atom doping
  in {2D} layered {MoS2} from a periodic table perspective},\ }\href
  {https://doi.org/10.1016/j.surfrep.2022.100567} {\bibfield  {journal}
  {\bibinfo  {journal} {Surface Science Reports}\ }\textbf {\bibinfo {volume}
  {77}},\ \bibinfo {pages} {100567} (\bibinfo {year} {2022})}\BibitemShut
  {NoStop}%
\bibitem [{\citenamefont {Qumar}\ \emph {et~al.}(2020)\citenamefont {Qumar},
  \citenamefont {Ikram}, \citenamefont {Imran}, \citenamefont {Haider},
  \citenamefont {Ul-Hamid}, \citenamefont {Haider}, \citenamefont {Riaz},\ and\
  \citenamefont {Ali}}]{Qumar2020}%
  \BibitemOpen
  \bibfield  {author} {\bibinfo {author} {\bibfnamefont {U.}~\bibnamefont
  {Qumar}}, \bibinfo {author} {\bibfnamefont {M.}~\bibnamefont {Ikram}},
  \bibinfo {author} {\bibfnamefont {M.}~\bibnamefont {Imran}}, \bibinfo
  {author} {\bibfnamefont {A.}~\bibnamefont {Haider}}, \bibinfo {author}
  {\bibfnamefont {A.}~\bibnamefont {Ul-Hamid}}, \bibinfo {author}
  {\bibfnamefont {J.}~\bibnamefont {Haider}}, \bibinfo {author} {\bibfnamefont
  {K.}~\bibnamefont {Riaz}},\ and\ \bibinfo {author} {\bibfnamefont
  {S.}~\bibnamefont {Ali}},\ }\bibfield  {title} {\bibinfo {title}
  {{Synergistic effect of Bi-doped exfoliated MoS2 nanosheets on their
  bactericidal and dye degradation potential}},\ }\href
  {https://doi.org/10.1039/d0dt00924e} {\bibfield  {journal} {\bibinfo
  {journal} {Dalton Transactions}\ }\textbf {\bibinfo {volume} {49}},\ \bibinfo
  {pages} {5362} (\bibinfo {year} {2020})}\BibitemShut {NoStop}%
\bibitem [{\citenamefont {Shibata}(1998)}]{Shibata_1998}%
  \BibitemOpen
  \bibfield  {author} {\bibinfo {author} {\bibfnamefont {H.}~\bibnamefont
  {Shibata}},\ }\bibfield  {title} {\bibinfo {title} {Negative thermal
  quenching curves in photoluminescence of solids},\ }\href
  {https://doi.org/10.1143/JJAP.37.550} {\bibfield  {journal} {\bibinfo
  {journal} {Japanese Journal of Applied Physics}\ }\textbf {\bibinfo {volume}
  {37}},\ \bibinfo {pages} {550} (\bibinfo {year} {1998})}\BibitemShut
  {NoStop}%
\bibitem [{\citenamefont {Mostaani}\ \emph {et~al.}(2017)\citenamefont
  {Mostaani}, \citenamefont {Szyniszewski}, \citenamefont {Price},
  \citenamefont {Maezono}, \citenamefont {Danovich}, \citenamefont {Hunt},
  \citenamefont {Drummond},\ and\ \citenamefont {Fal'ko}}]{Mostaani2017}%
  \BibitemOpen
  \bibfield  {author} {\bibinfo {author} {\bibfnamefont {E.}~\bibnamefont
  {Mostaani}}, \bibinfo {author} {\bibfnamefont {M.}~\bibnamefont
  {Szyniszewski}}, \bibinfo {author} {\bibfnamefont {C.~H.}\ \bibnamefont
  {Price}}, \bibinfo {author} {\bibfnamefont {R.}~\bibnamefont {Maezono}},
  \bibinfo {author} {\bibfnamefont {M.}~\bibnamefont {Danovich}}, \bibinfo
  {author} {\bibfnamefont {R.~J.}\ \bibnamefont {Hunt}}, \bibinfo {author}
  {\bibfnamefont {N.~D.}\ \bibnamefont {Drummond}},\ and\ \bibinfo {author}
  {\bibfnamefont {V.~I.}\ \bibnamefont {Fal'ko}},\ }\bibfield  {title}
  {\bibinfo {title} {Diffusion quantum monte carlo study of excitonic complexes
  in two-dimensional transition-metal dichalcogenides},\ }\href
  {https://doi.org/10.1103/PhysRevB.96.075431} {\bibfield  {journal} {\bibinfo
  {journal} {Phys. Rev. B}\ }\textbf {\bibinfo {volume} {96}},\ \bibinfo
  {pages} {075431} (\bibinfo {year} {2017})}\BibitemShut {NoStop}%
\end{thebibliography}%
\ifarXiv


\section*{Supplementary material}

Supplementary Material for:

The impact of electrical contacts on the optical properties of a MoS$_2$ monolayer

Agata Zielińska,[1, a)] Mateusz Dyksik,[1] Alessandro Surrente,[1] Jonathan Eroms,[2] Dieter Weiss,[2] Paulina Płochocka,[1, 3] and Mariusz Ciorga[1]

[1)]*Department of Experimental Physics, Wrocław University of Science and Technology, Wybrzeże Wyspiańskiego 27, 50-370, Wrocław, Poland*

[2)]*Institute for Experimental and Applied Physics, University of Regensburg, Universitätsstraße 31, 93053, Regensburg, Germany*

[3)]*Laboratoire National des Champs Magnétiques Intenses, 143 avenue de Rangueil, 31400, Tolouse, France*

(Dated: 25 April 2025)

[a)]agata.zielinska@pwr.edu.pl

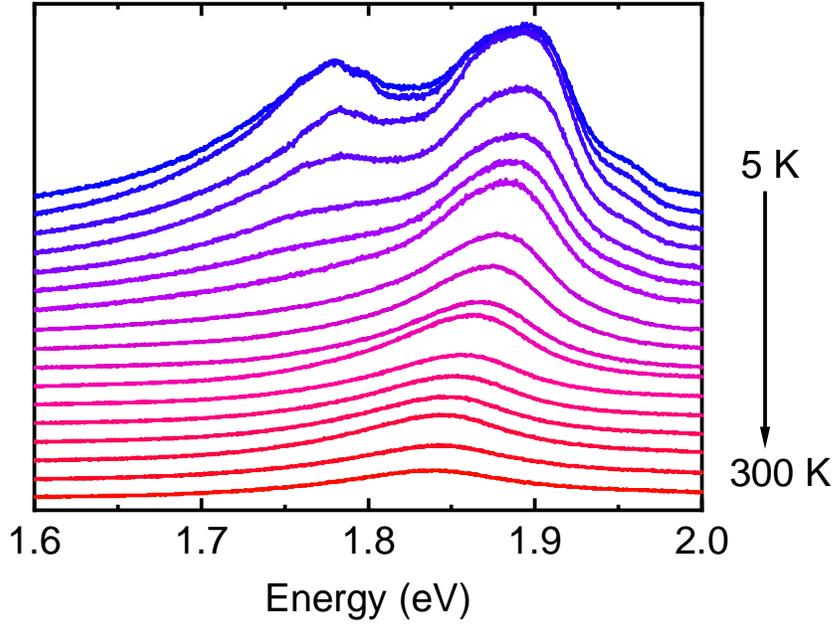

FIG. S1: Temperature series of sample A measured in range 5–300 K with a 20 K step. We can observe that the defect D is suppresed with increasing the temperature.

In Fig. S1. we show the temperature dependence of the PL emission from sample B. Figure 4 in the main text presents the corresponding Arrhenius plots of PL intensity of L and D lines. L line was fitted with a function containing two activation energies for PL quench, as only one activation energy was not enough to reproduce the experimental data. We obtain two activation energies of 4.6(20) meV and 37(14) meV. D line was fitted with a function containing three activation energies - one for the increase of PL and two for the PL quench. It was nessesary to use three activation energies to correctly reproduce the data. We obtained 2.5(8) meV activation energy for the PL increase, and 5.2(5) meV and 46(10) meV for the two processes of PL quench. We compared the obtained values with the binding energies of various excitonic complexes on different types of defects in transition-metal dichalcogenides,calculated with Monte Carlo simulations performed by Mostaani et al.[1]. The first activation energy for L line could correspond to an exciton bound on donor ($D^+X$) or acceptor ($A^-X$) in the lattice structure. Their binding energies are equal to 7.2 meV and 2.7 meV, respectively. Taking into account the uncertainty of obtained values, we can conclude that the first activation energy corresponds probably to the process of $A^-X$ dissociation. In case of $MoS_2$ it could be an exciton bound to a carbon atom placed in a sulfur vacancy[2]. The second activation energy equal to 37 meV can correspond to a negative trion bound to a donor

($D^0X$) or positive trion bound to an acceptor ($A^0X$). Their binding energies are equal to 32.4 meV and 31.7 meV, respectively. Taking into account the fact that MoS$_2$ is naturally n-doped, $D^0X$ is more probable. In our sample it would mean a negative trion bound to a hydrogenized sulphur vacancy[3]. Two excitonic complexes emitting near 1.8 eV would be in agreement with a PL shape which is very broad and looks like more than one Gaussian. In case of line D we observe an initial PL increase with an activation energy of 2.5 meV, which can correspond to $A^-X$ dissociation. It could mean that the carriers previously dissociated from the line L are then bound to the states responsible for the line D emission state, nincresing the P intensity. The additional. carriers can also come from a different reservoir, which is not described here. The first activation energy of the PL quench 5.2 meV may correspond to dissociation of $D^+X$ or $A^-X$. The second activation energy equal to 46 meV may correspond to a biexciton bound to acceptor, which activation energy is equal to 48 meV. If we assume the diffusion of Bi atoms during the evaporation process and substitution of sulphur atoms, then Bi atoms would serve as acceptors, as bismuth has one valence electron less than sulphur Taking this into account, we can conclude that the first activation energy for PL quench probably corresponds to a dissociation of an exciton bound to Bi atom acceptor ($A^-X$) and the second activation energy corresponds to a biexciton bound to Bi atom acceptor. The fact that we do not observe the free biexciton emission could mean that our sample structure has so many defects, that all biexcitons are bound to them and there are no free biexcitons. The majority of excitons and trions could also be bound to defects and it would be consistent with the very low PL emission of the free complexes.

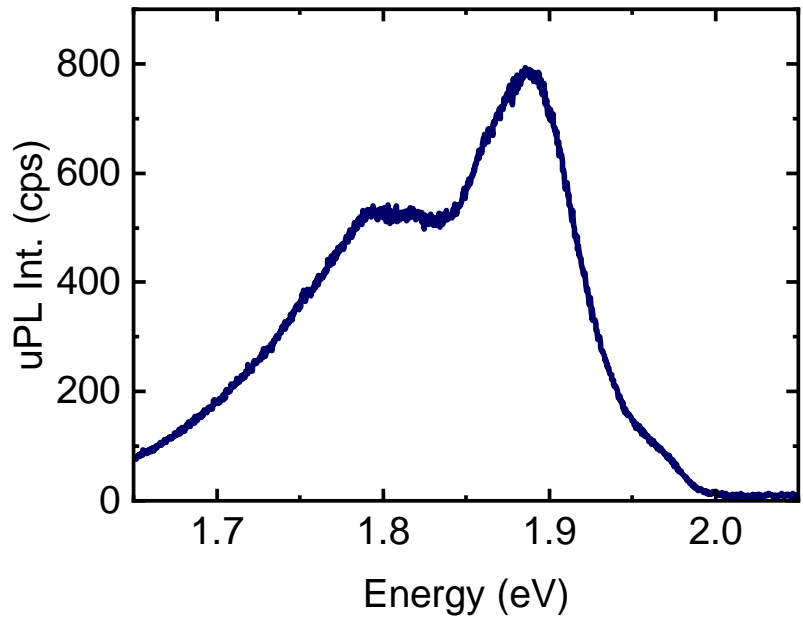

FIG. S2: μPL of the second sample with Bi contacts.

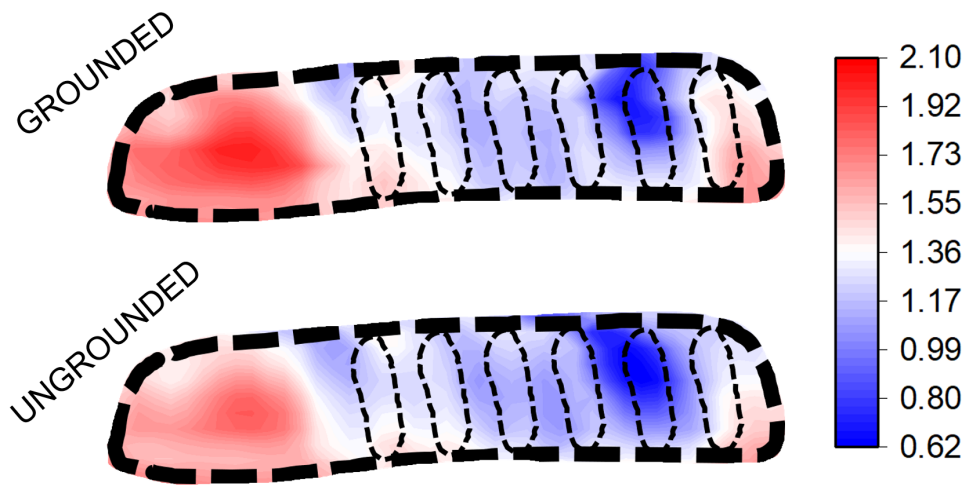

FIG. S3: Trion to defect ratio of sample B at 50 K. The differences between grounded and ungrounded contacts at both temperatures are much less visible than for sample A.

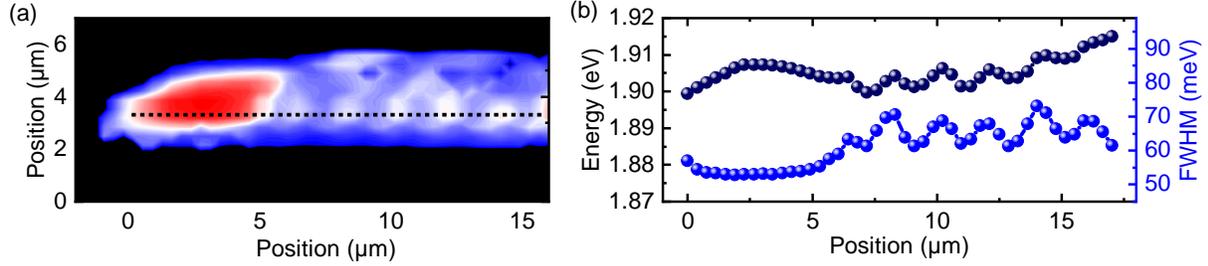

FIG. S4: (a) µPL map obtained by integrating the spectral range corresponding to the trion and neutral exciton emission (1.86–2 eV). The bright parts in the middle of the map denote the positions of the contacts. (b) Position dependence of the energy centroid (center of mass) and the full width at half maximum (FWHM) of the integrated peak extracted across the dashed line in (a). Parts of the flake laying on the contacts characterize with higher intensity, smaller centroid and smaller FWHM of the integrated emission. The emission from the left part of the flake (far away from the contacts) is characterized by much smaller broadening than the parts on or between contacts. It means that the emission from the flake around contacts is inhomogeneous, which is probably also a consequence of the strain induced during the flake transfer on the contacts.

**Fabrication of electrical contacts** - Silicon substrates with 285 nm $SiO_2$ layer were first cleaned in $O_2$ plasma for 5 minutes. Then, the substrates were spin-coated twice with PMMA layers (200k 9% and 950k 5%) and baked for 5 minutes at 150 °C after each process. Subsequently, the contacts pattern was written on each substrate by electron-beam litoghaphy with using self-prepared mask in a Zeiss Auriga scanning electron microscope. Small parts of the contacts were written with 400 µC/cm$^2$ area dose and 20 µm aperture. Bigger parts were written with 350 µC/cm$^2$ area dose and 120 µm aperture. The pattern was then developed for 20 s in methyl isobutyl ketone (MIBK) diluted in isopropanol with 1:3 ratio and cleaned in isopropanol. Afterwards, the substrates were put in an UHV system where 5 nm of titanium and 20 nm of gold or 20 nm of bismuth and 80 nm of gold were evaporated. The PMMA resist was finally resolved in hot acetone at 60 °C.

**Exfoliation and transfer** – $MoS_2$ flakes were produced by means of mechanical exfoliation from bulk crystals (HQ graphene) and transferred onto Polydimethylsiloxane (PDMS) gel-films. Then, an optical microscope was used to find monolayers by the optical contrast. Finally, the monolayers were dry-transferred onto substrates with previously evaporated gold contacts (substrates were heated to 80 °C).

**AFM ironing** – MoS$_2$ flake on each gold contact was one-way-scanned by Park AFM in contact mode with a non-contact cantilever. The ironing force was set to 100 nN, the scan rate was set to 0.6 Hz. The tip velocity was $\sim 5\,\mu$m/s and the resolution was set on 512 px, which gives 1 nm to 2 nm between each line scan.

**Transport measurements** - I-V characteristics were measured in a wide temperature range in a superconducting magnet system (Oxford Instruments). Yokogawa 7651 was used as a voltage source, and Aglient 34410A multimeter with Ithaco 1211 preamplifier were used to measure drain-source voltage. Keithley SMU 2400 source meter was used to set backgate voltages.

**Optical spectroscopy** - Measurements were performed in an optical helium flow cryostat (Oxford Instruments). The excitation source for microphotoluminescence was a 532 nm continuous wave laser. The white light source for microreflectance was a Tungsten-Halogen Light Source (Thorlabs), which was spatiallz filtered through a 50 μm pinhole, which yields the final spot size of $\sim 4\,\mu$m. The laser light and the white light were focused on the sample surface by 0.75 NA and 50× microscope objective (Mitutoyo). The same objective was used to collect the output signal, which was then directed to the spectrometer (500 mm focal length, 300 groove/mm grating and 150 μm slit width) equipped with a liquid nitrogen cooled CCD camera (Princeton Instruments).

## REFERENCES

[1] E. Mostaani, M. Szyniszewski, C. H. Price, R. Maezono, M. Danovich, R. J. Hunt, N. D. Drummond, V. I. Fal'ko, "Diffusion quantum Monte Carlo study of excitonic complexes in two-dimensional transition-metal dichalcogenides," *Physical Review B*, 075431, 2017.

[2] Park, Y., Li, N., Jung, D. et al. "Unveiling the origin of n-type doping of natural MoS2: carbon," *npj 2D Mater Appl* 7, 60, 2023

[3] Singh, Akash and Singh, Abhishek Kumar, "Origin of *n*-type conductivity of monolayer MoS$_2$," *Phys. Rev. B* 99, 12, 2019



\end{document}
%